# Trustworthy AI and Robotics and the Implications for the AEC Industry: A Systematic Literature Review and Future Potentials

**Newsha Emaminejad**[1] **and Reza Akhavian, Ph.D.**[2*]

[1]Graduate Student, Department of Civil, Construction, and Environmental Engineering, San Diego State University, 5500 Campanile Dr., San Diego, CA 92182; e-mail: nemaminejad859@sdsu.edu

[2]Assistant Professor, Department of Civil, Construction, and Environmental Engineering, San Diego State University, 5500 Campanile Dr., San Diego, CA 92182; e-mail: rakhavian@sdsu.edu

* Corresponding Author

## Abstract

Human-technology interaction deals with trust as an inevitable requirement for user acceptance. As the applications of artificial intelligence (AI) and robotics emerge and with their ever-growing socio-economic influence in various fields of research and practice, there is an imminent need to study trust in such systems. With the opaque work mechanism of AI-based systems and the prospect of intelligent robots as workers' companions, context-specific interdisciplinary studies on trust are key in increasing their adoption. Through a thorough systematic literature review on (1) trust in AI and robotics (AIR) and (2) AIR applications in the architecture, engineering, and construction (AEC) industry, this study identifies common trust dimensions in the literature and uses them to organize the paper. Furthermore, the connections of the identified dimensions to the existing and potential AEC applications are determined and discussed. Finally, major future directions on trustworthy AI and robotics in AEC research and practice are outlined.

# 1. Introduction

AI and robotics (hereinafter AIR) will soon be able to affect various aspects of human life. As the role of AI in our lives becomes more prominent and with the emergence of autonomous agents in industrial and social settings around the world, the need to build trust between these agents and their human counterparts is becoming ever more apparent (Kessler, Larios et al. 2017). The National Academies of Science, Engineering, and Medicine and the Royal Society identified trust, transparency, and interpretability as some of the key socio-technical challenges in designing, evaluating, and deploying AI systems (National Academies of Sciences and Medicine 2018). Research has shown that an explainable and transparent decision-making process can help users gain an appropriate level of trust in intelligent agents (Wang, Pynadath et al. 2016). However, the algorithmic complexity and abstraction of such a decision-making process make it a challenging task to provide straightforward and comprehensible explanations (Glikson and Woolley 2020). Additionally, recent studies indicate that while transparency is a key factor, it is not the only factor that affects trust in AI-powered systems (Winfield and Jirotka 2018).

In the context of AIR-enabled systems, it has been shown that new technology is more accepted and trusted when it behaves and functions as we expect other humans to do (Nomura 2017). Furthermore, human biases and stereotypical concepts can also influence the level of trust in such systems. For example, AIR agents with a male voice are expected to be associated with more technical tasks, while those with female voices are assumed to be related to household or lifestyle-related tasks (Chemaly 2016). This and other examples in the literature show that besides developing transparency and positive perception (i.e., technical aspects of technology adoption), exploring the factors that deal with the psychological side of the issue is equally important.

Over the last decade, the architecture, engineering, and construction (AEC) industry has shown significant potential to be disrupted by AIR (Wisskirchen, Biacabe et al. 2017, Jose, Steffen et al. 2018, Delgado, Oyedele et al. 2019, Sumana 2019). Nevertheless, AEC practitioners are generally reluctant to adopt new technologies and the use of antiquated work processes is prevalent in the industry (Bock 2015). Additionally, small businesses comprise the vast majority of the industry with a share of 82.3% (Kobe 2002) and smaller companies are known to be often the late majority and laggards in technology adoption (Peltier, Zhao et al. 2012). Research has shown that building a culture of trust in the potential and reliability of AIR systems can play a key role in enhancing adoption levels within the AEC industry (Schia 2019). Considering the myriad of parameters that affect trust, the opacity of AIR systems working mechanism, and the complex and unique nature of AEC projects, this paper presents a comprehensive review of the literature on trustworthy AIR and AIR applications in the AEC and illustrates future research directions and potentials in the area of trustworthy AIR in AEC. Figure 1 shows the paper's scope.

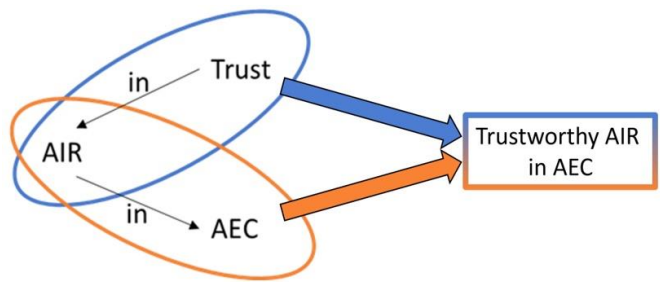

**Figure 1. The interconnection of the application of AIR in the AEC industry and the role of trust in using AIR as the scope of this study**

The remainder of the paper is organized as follows. First, an overview of the research background including studies on trust in technology and the AIR adoption in AEC is provided. Next, trust dimensions used to organize this review are introduced and defined, and following that, the systematic review approach is described. Finally, directions for future research are proposed and a conclusion is presented.

## 2. Research Background
### 2.1. Trust in Technology

To explain people's attitude toward a newly introduced technology, a widely-used model called the Technology Acceptance Model (TAM) has emerged in the information systems domain with applications in various fields such as AIR (Venkatesh and Davis 2000, Lee, Kozar et al. 2003, King and He 2006, Sepasgozaar, Shirowzhan et al. 2017). TAM explains that the most important parameters involved in acceptance of and trust toward a new technology are the perceptions of a) usefulness and b) ease of use. For example, being benevolent (or not threatening) and conformed to our ideas influenced by societal contexts can make new technology more acceptable and trustworthy (Wu, Zhao et al. 2011). However, TAM fails to account for factors such as transparency, reliability, and privacy that may influence trust between humans and embodied intelligence (Alhashmi, Salloum et al. 2019). Therefore, a significant number of research studies have focused on these trust parameters as adoption drivers. A detailed description of these parameters is provided in Section 3 and relevant research studies are reviewed in Section 5.

In this research, AIR is regarded as an interdisciplinary science encompassing fields including computer science, engineering, mathematics and statistics, and psychology that enable learning from existing data and past experience to perform tasks that typically require humans' intellectual processes (Hild and Stemmer 2007, Kotu and Deshpande 2018, Glikson and Woolley 2020). AIR applications studied in this research include embodied intelligence (e.g., a robot) and software or hardware that can be used stand-alone or distributed through computer networks.

### 2.2. AIR Adoption in the AEC Industry

Technical processes and workflows in the AEC industry are often infamous for inefficiency, safety hazards, and workforce issues (Hallowell 2010). In other industries such as automotive, health care, and manufacturing, AIR-enabled systems helped alleviate such challenges (Pagliarini and Lund 2017, Pillai, Sivathanu et al. 2021). Digital transformation including the use of AIR is set to revolutionize the AEC industry as well (Darko, Chan et al. 2020). Many researchers have studied and explained the potential improvements made by AIR-based technologies for different phases of construction projects (Stepanov and Gridchin 2018, Xiao, Liu et al. 2018, Aripin, Zawawi et al. 2019, Delgado, Oyedele et al. 2019, Yahya, Hui et al. 2019, Elhouar, Hochscheid et al. 2020). Nevertheless, a one-size-fits-all study of technology adoption does not do justice in the AEC field due to the multifaceted nature of construction projects involving multiple stakeholders, decentralized production, extensive use of subcontractors, lack of standard workflows, and various project types. Therefore, researchers have used various classification schemes to categorize technology adoption approaches, drivers, or barriers in the AEC fields. For example, Hatami et al. (2019) focused on construction manufacturing systems and categorized the use of AI in this area into four main applications of planning and design, safety of autonomous equipment, and monitoring and maintenance (Hatami, Flood et al. 2019). In another study, a systems-framework was presented by Pan et al. (2020) to identify key factors influencing future implementation of construction robotics in Hong Kong. They identified eleven interrelated factors, critical for shaping the future trajectory of construction robotic applications, including construction costs, government support, and the scale of prefabrication, and suggested that implementation of robots in construction suffers from lack of interdisciplinary and non-technical studies (Pan, Linner et al. 2020). The four main factors hindering the adoption and acceptance of robotics and automated systems in the construction industry were identified in another study as (1) contractor-based financial factors, (2) client-based financial factors, (3) technical and work-culture factors, and (4) weak business case factors (Delgado, Oyedele et al. 2019). In another scientometric study and using a science-mapping method, it was concluded that genetic algorithms, neural networks, fuzzy logic, fuzzy sets, and machine learning have been the most widely used AI methods in AEC to address topics and issues such as optimization, simulation, uncertainty, project management, and bridges (Darko, Chan et al. 2020). Sacks et al. (2020) reported AI fields such as machine learning, deep learning, and computer vision as top AEC-supporting technologies besides more established technologies such as BIM, 3D printing, and Simultaneous Localization and Mapping (SLAM).

With respect to adoption driver and barriers, research has identified three key parameters of (1) promotion and support of research and evolution of supporting technologies, (2) simultaneous or non-simultaneous developments in various fields of application, and (3) inter-disciplinary collaboration between AEC industry and technology companies as key drivers of Research and Development (R&D) in construction robotics (Basaif, Alashwal et al. 2020). In another recent study, Sacks et al (2020) concluded that lack of complete and accessible information in models is a major hindrance to broad adoption of BIM and AI technologies in the construction industry that prevents effective exploitation of the information they provide (Sacks, Girolami et al. 2020).

## 3. Trust Dimensions

State-of-the-art research has used different taxonomic categorizations to classify factors influencing trust in AIR systems (Glikson and Woolley 2020, Kok and Soh 2020, Sanneman and Shah 2020, Wing 2020). A notable study in this area suggests that different trust dimensions affect human trust regardless of the form of AI representation (robot, virtual, and embedded) and thus an integrated review approach can provide a comprehensive view of trust in AI within a specific disciplinary field (Glikson and Woolley 2020). To identify the key trust dimensions most applicable to AEC projects, applications, and project types and phases, the authors have previously conducted a thematic analysis (Emaminejad, North et al. 2021). Inspired by the findings of that study and the literature on trustworthy AIR, in this research common trust dimensions are used to organize a comprehensive and systematic review of the literature. The eight dimensions identified are Transparency, Interpretability, Reliability, Safety, Performance, Robustness, Privacy, and Security. Considering semantic similarities and in order to create a more focused discussion for AEC applications, the eight dimensions are presented in pairs for the purpose of this review. Therefore, four key trust dimensions are used to review empirical research on trustworthy AIR and implications for AEC research. Figure 2 illustrates the four trust dimensions used to organize this paper.

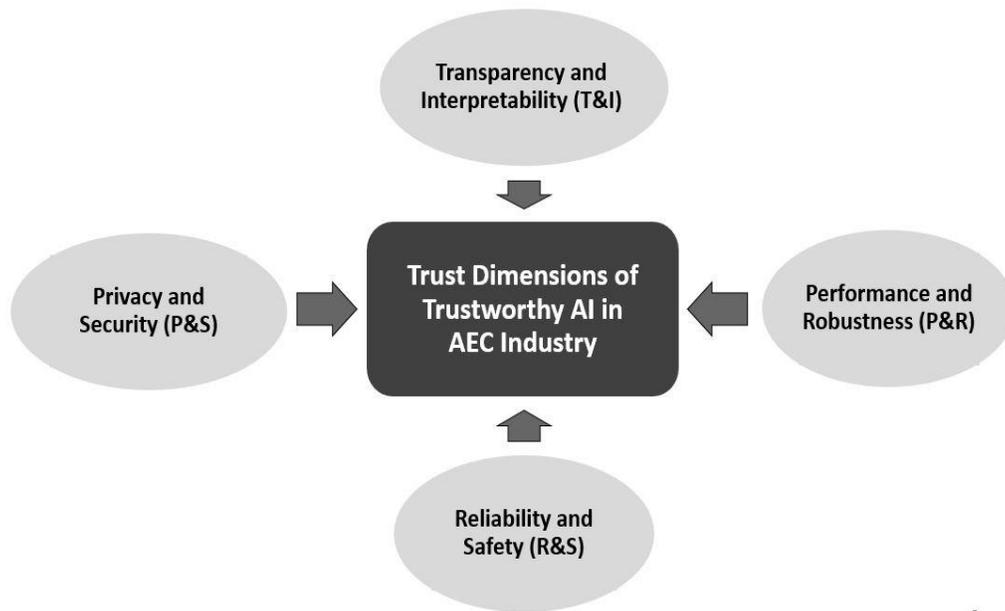

**Figure 2. Selected trust dimensions for a systematic review of trustworthy AIR in the AEC industry**

### 3.1. Transparency and Interpretability (T&I)

In AIR applications, transparency is often related to the concept of interpretability where the operations of a system can be understood by a human through introspection or explanation (Biran and Cotton 2017, Haibe-Kains, Adam et al. 2020). In computer science, explainable AI (xAI) is a trending research topic (Fox, Long et al. 2017). In construction, it is especially important since technologies are often not developed in-house. Therefore, the end-users may need to understand what happens between the input and output (i.e., transparency) and the output should be understandable and meaningful to them (i.e., interpretability).

### 3.2. Reliability and Safety (R&S)

Reliability is concerned with the capacity of the models to avoid failures or malfunctions and exhibit the same and expected behavior over time (Hoff and Bashir 2015). It can be seen from both technical and cognitive/psychological (e.g., affect) perspectives (Hamon, Junklewitz et al. 2020). As a related dimension, and in the context of adopting AIR in AEC, safety also has to be studied from both technical (i.e., avoiding accidents with heavy-duty construction robots) and psychological (i.e., AIR applications seen as job companions, not competitors) (Dahlin 2019, Ivanov, Kuyumdzhiev et al. 2020).

### 3.3. Performance and Robustness (P&R)

Ability, accuracy, or competence are similar concepts used in the AIR literature to refer to the importance of performance indicators in gaining human trust (Hoffman, Johnson et al. 2013, Siau and Wang 2018, Wing 2020). Performance plays a key role in trust trajectories as research has shown that human develops a performance expectation toward AIR systems and failing to satisfy that expectation (or exceeding it) after interaction has major effect on trust (Robinette, Howard et al. 2017, Glikson and Woolley 2020). Robustness, as an important trust dimension, refers to consistency of performance in different situations (Wing 2020). This is particularly important in AEC applications as the environment and workflows are ever-changing within and between projects and failure to adapt to the new setting would result in reduced levels of trust.

## 3.4. Privacy and Security (P&S)

Humans' trust in technology is highly influenced by the levels of P&S involved in technology implementation (Hamon, Junklewitz et al. 2020). Both dimensions fall under the category of ethical or responsible AI paradigm where the protection human identity and sensitive data and vulnerability (i.e., security) of the system toward attacks that breach privacy are known as leading trust drivers (Orr and Davis 2020). In the AEC literature, workers and client's sensitive data vulnerable to privacy leaks (Choi, Hwang et al. 2017) as well as cybersecurity issues pertaining to the use of cameras, drones, and other sensing devices have been widely regarded as barriers to technology adoption (Choi, Hwang et al. 2017, Kozlovska, Klosova et al. 2021). P&S issues in AEC can be exacerbated by the growth of automated and intelligent agents and have to be addressed in the context of AIR-based systems.

## 4. Systematic Review Methodology

In this study, a systematic literature review (SLR) approach developed by Lockwood and Oh (2017) was used to illustrate a comprehensive picture of the literature and review the most relevant publications. Figure 3 presents a schematic overview of the review approach adopted in this research.

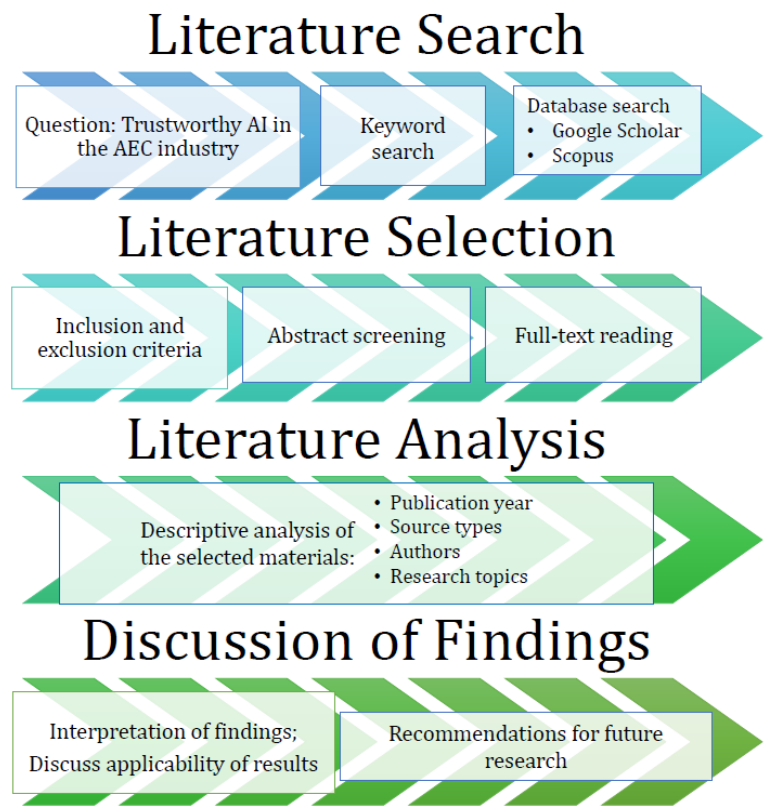

**Figure 3. Schematic overview of the systematic review approach used in this research study, comprising literature search, selection, and analysis and also analysis of research findings**

In Stage 1 of the SLR, a comprehensive search was conducted using Scopus and Google Scholar databases to find articles published in peer-reviewed journals and conference proceedings. The search keywords included "Artificial Intelligence applications in construction management", "AI applications in construction management", "Robotics in construction management", "Construction Automation", "Trust in AI", "Trustworthy AI", "Acceptance modeling for AI", "Ethical AI", "Transparent and explainable AI" and "Reliable and safe AI". In total, 520 publications were collected for review and imported to a reference managing software EndNote (Agrawal and Rasouli 2019). In Stage 2, abstracts, highlights, and key words were retrieved, analyzed, and reviewed, and those studies that met the inclusion criteria were reviewed in full. The inclusion criteria were to use only peer-reviewed papers; publication date must be after 2010 (papers published prior to 2010 were not included due to the fast pace of AIR research advancement in the last two decades); and the focus of the paper must be on trust (or its dimensions) in AIR or application of AIR in AEC. Also, papers that focused solely on automation or robotics in construction without an AI component were excluded from the database. This literature selection and screening process reduced the number of total reviewed articles to 484. The majority of AI in the AEC subset of these papers are published in journal venues including Automation in Construction, Journal of Computing in Civil Engineering, Advanced Engineering Informatics, Journal of Construction Engineering and Management, Journal of Information Technology in Construction, and Computer-Aided Civil and Infrastructure Engineering. Stage 3 involved creating a spreadsheet describing these papers (i.e., title, authors, publication year, and keywords) and a detailed review of the content. A summary of the papers with their publication year and main focus is presented in Table 1 and a keyword co-occurrence map of these papers created by the VOSviewer software (Van Eck and Waltman 2013) is shown in Figure 4. The size of the nodes is proportionate with the frequency of the keyword occurrence and the distance between two nodes is inversely proportionate with the strength of the relation between the keywords in the literature analyzed. Additionally, the nodes and links are color-coded to reflect the publication age. Finally, Stage 4 of the SLR is described in Sections 5 and 6 of this paper.

**Table 1. Classification of the reviewed papers based on trust parameters.**

| Topics | Trust in AIR | | | | AI in AEC | Total |
|---|---|---|---|---|---|---|
| | *T&I* | *R&S* | *P&R* | *P&S* | | |
| **Number of Papers** | 143 | 82 | 66 | 76 | 212 | 484 |

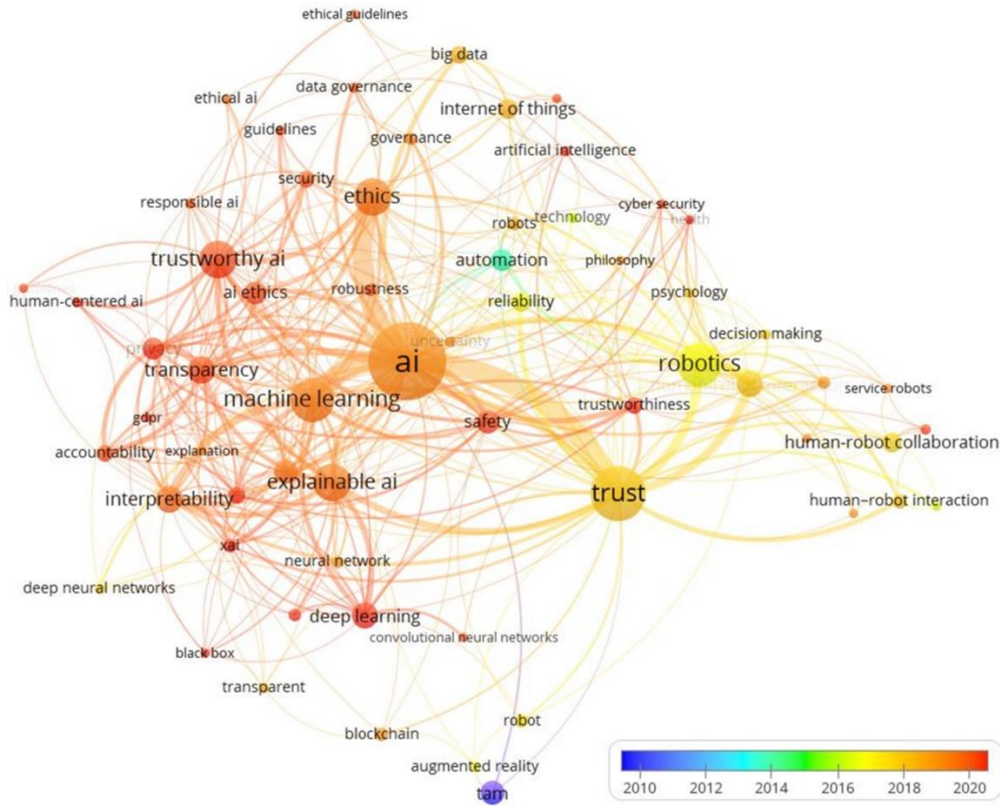

**Figure 4. Map of Keyword Co-Occurrences**

## 5. Trust in AIR and its Implications for AEC Adoption

Trust is defined as "*the attitude that an agent will help achieve an individual's goals in a situation characterized by uncertainty and vulnerability*" (Lee and See 2004), and as "*the reliance by an agent that actions prejudicial to their well-being will not be undertaken by influential others*" (Hancock, Billings et al. 2011). In this study, efforts to gain higher levels of trust in AI/robotics systems are reviewed in three Classes of Transparency and Interpretability, Reliability and Safety, Robustness and Accuracy, and Privacy and Ethics.

### 5.1. Transparency and Interpretability (T&I)

As suggested by the panel on the "One Hundred Year Study on Artificial Intelligence, Artificial Intelligence and Life in 2030," (Peter Stone, Rodney Brooks et al. 2016) "well-deployed artificial intelligence prediction tools have the potential to provide new kinds of transparency about data and inferences, and may be applied to detect, remove, or reduce human bias, rather than reinforcing it." Technologies that involve obscure processes between the input fed to the system and the generated output are generally less trustworthy than those with transparent and interpretable inner-workings (Carvalho, Pereira et al. 2019). The lack of a common language is a key barrier to achieving transparency in AI as varying dentitions exist across disciplines (National Academies of Sciences and Medicine 2018). "Openness" and "explainability" are two neighboring terms that are commonly used in the literature to refer to the T&I concept in AI systems and address the notions implied by "black box" terminology (Larsson and Heintz 2020). In particular, the concept xAI has recently gained tremendous momentum in computer science research (Ribeiro, Singh et al. 2016, Biran and Cotton 2017, Lepri, Oliver et al. 2018, HLEG 2019). This research trend is motivated by intriguing results of deep learning (e.g., mistakenly labeling a tomato as a dog (Szegedy, Zaremba et al. 2013)) to provide an overview on T&I of data-driven algorithms (Biran and Cotton 2017, Doran, Schulz et al. 2017, Miller, Howe et al. 2017, Samek, Wiegand et al. 2017, Guidotti, Monreale et al. 2018). Particularly in robotics, explainable agents are those autonomous agents capable of providing explanations about their actions and the rationale behind their decisions (Langley, Meadows et al. 2017). Understanding the behaviors of robots that results in understanding their actions is known as mentalizing or mindreading which is derived from the Theory of Mind (ToM). According to the ToM, that humans estimate the actions of other humans by observing their behavior and attributing mental states thereby attempting to understand their own perspective (Goldman 2012, Hellström and Bensch 2018). Research studies in the field of human-robot interaction (HRI) have confirmed that humans show a similar trait (i.e., applying ToM) to non-human objects and robots (Lee, Lau et al. 2005, Hellström and Bensch 2018). More generally, and to gauge humans' trust in applied AI, it is important to account for how ordinary users understand explanations and assess their interactions with the system (e.g., tool, service) (Miller 2017). Developing higher extents of T&I equips users with higher levels of understanding of and thereby trusting in the intelligent agents or predictive models (Mercado, Rupp et al. 2016, Kopitar, Cilar et al. 2019, Marcos, Juarez et al. 2020).

*5.1.1. T&I in AEC AIR*

In industrial applications, T&I can be viewed as the most important factor in determining the level of trust in AI and robotics systems given the fact that such systems are often designed and developed by companies external to the end-user organization (Desouza, Dawson et al. 2020). This is particularly true in construction companies where technologies are not developed in-house and are purchased or licensed from outside developers and the obscure nature of decisions made by AI-based systems make practitioners hesitant to trust them. For example, sustainable design can benefit from AI adoption where intelligent agents offer choices for energy efficiency and green materials, but unless there is human intelligence into why this decision was made, the sustainable choices are not trustworthy (Gilner, Galuszka et al. 2019). From the HRI standpoint, T&I in construction robotics can be achieved if the robot is flexible in the type and extent of help it offers. The level of collaboration may differ as some tasks are structured and specified to be completely performed by a robot, while some other tasks may require human engagement (i.e., human-in-the-loop or HITL) (Brosque, Galbally et al. 2020). In construction, HITL applications comprise the majority of the current applications. In one example, and with an emphasis on the unstructured, dynamic job site environment and conditions, Follini et al. (2020) proposed a collaborative robotic platform focused on HILT applications. Their proposed platform is programmed to follow the operator while it carries heavy materials and equipment and stops when its sensors detect the close distance to the operator or any obstacles. Additionally, the robot can navigate through the site based on known geometric and semantic information of the building project by connecting the Robot Operating System (ROS) and the Building Information Model (BIM) (Follini, Terzer et al. 2020).

Very little is known regarding T&I in AI-based construction application. Researchers have examined the potential of Blockchain technology to increase transparency and simplify interpretably in different phases of construction (Turk and Klinc 2017, Pan and Zhang 2021). Blockchain technology promotes peer-to-peer digital transaction management which can be used as a secure and transparent method among users. In simple words, Blockchain can add value to AI by explaining AI decisions, mitigating disastrous risks, increasing AI efficiency, and improving data accessibility and decentralization (Nassar, Salah et al. 2020). In another study, a framework was proposed based on Blockchain technology to facilitate the use of BIM, and improve data transparency and security in construction projects (Singh and Ashuri 2019). Furthermore, use of Blockchain in AEC industry and its current challenges was studied by Wang et al. They proposed 3 Blockchain-based applications for construction projects, namely Blockchain-based contract management, Blockchain-based supply chain management, and Blockchain-based equipment leasing which can be utilized by AEC professionals (Wang, Wu et al. 2017).

**5.2. Reliability and Safety (R&S)**

R&S are two cornerstones of trust in AI-powered systems (Marcus and Davis 2019). Both concepts are related to trust from a performance rather than a moral angle (Malle and Ullman 2021). They are two interconnected paradigms in the context of trust between the human user and robot agent. On the one hand, safe interaction with robots make them more reliable and thus trustworthy (Murashov, Hearl et al. 2016). On the other hand, increased reliability may lead to overtrust and higher potential complacency when the robots act unsafely (Lyons 2013, Merritt, Heimbaugh et al. 2013). This tradeoff highlights the importance of a process called trust calibration within a human-robot team to achieve an appropriate level of trust given both human and robot capabilities. Trust calibration process enables a human to accurately recognize the risks associated with trusting a person or machine (Lee and See 2004). In this process, the human user learns the abilities, reliability, and failure modes of the agent to avoid overtrust and undertrust (Parasuraman, Molloy et al. 1993). If the user undertrusts the intelligent agent, trust calibration helps them to gradually build trust in the system with experience and iterative interactions (King-Casas, Tomlin et al. 2005). On the other hand, in an overtrust situation, the user accepts too much risks as they believe that the system will mitigate those risks (Borenstein, Wagner et al. 2018).

Another tradeoff that affects trust from an R&S perspective, is the amount of human effort versus robot autonomy. Reliability is an HRI variable that needs to be evaluated as a function of robot autonomy. Beer et al. (2014) discussed this concept by proposing levels of robot autonomy (LORA) for HRI (Beer, Fisk et al. 2014). Their proposed LORA classification starts at a stage where the human is in full control and the process is Manual and continues with similar approximations of a robot's autonomy along a continuum that ends with Full Autonomy as illustrated in Figure 5.

All these classifications are analyzed based on the human, robot, or both engagement in sensing, planning, and acting. However, LORA and similar widely influential taxonomies of automation levels such as the classic Sheridan-Verplank levels of automation (Sheridan and Verplank 1978) and recent classification schemes for self-driving cars (SAE 2014, Brooks 2017) levels of autonomy are criticized due to the fact that they only represent situations in which boosted levels of automation will certainly result in restricted and limited human control (i.e., one-dimensional levels of automation). In an effort to address these constraints, Shneiderman (2020) proposes the Human-Centered Artificial Intelligence (HCAI) framework to produce designs that are reliable, safe, and trustworthy. HCAI indicates that despite what one-dimensional taxonomies suggest, achieving high levels of human control, and at the same time, high levels of automation (i.e., two-dimensional HCAI) is possible and in fact, leads to increased human performance (Shneiderman 2020). This two-dimensional HCAI toward reliable, safe, and trustworthy AI is shown in Figure 6.

| 9 | Full Autonomy | Robot performs all aspects of a task autonomously |
| 8 | Supervisory Control | Robot does all aspects of task, but human continuously monitors the process. Human can override and set new goals |
| 7 | Executive Control | Human sets an abstract high-level goal. Robot autonomously senses environment, sets the plan, and takes action. |
| 6 | Shared Control with Robot Initiative | Robot does all aspects of the task and can ask human for assistance in setting new goals and plans, if there are difficulties and ahcllenges. |
| 5 | Shared Control with Human Initiative | Robot autonomously senses the environment, develops plans and goals, and takes actions. Human monitors the process and may intervene and set new goals and plans if the robot is facing challenges. |
| 4 | Decision Support | Both the human and robot sense the environment and generate a task plan. Human makes the final choice and commands robot to execute tasks. |
| 3 | Batch Processing | Both the human and robot monitor and sense the environment. Only human sets the goals and plans for the task. Robot executes the assigned tasks. |
| 2 | Assisted Teleoperation | Robot senses the environment and chooses to intervene with task that human assist with all aspects of it. |
| 1 | Tele-operation | Human is responsible for sensing and planning. Robot assists the human with task execution. |
| 0 | Manual | Human does all aspects of the task with no involvement of robot. |

**Figure 5. One-Dimensional levels of robot involvement in operations (modified from Beer 2014)**

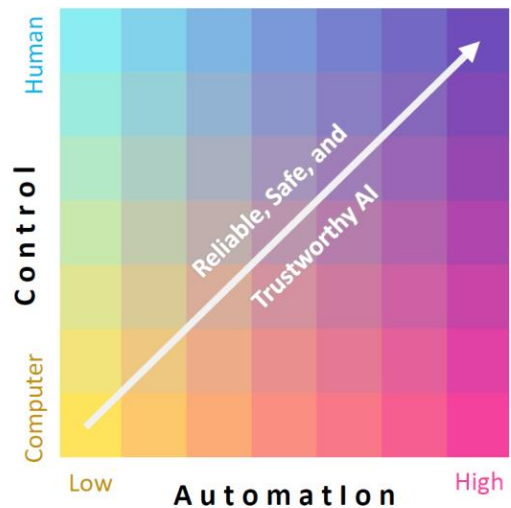

**Figure 6. Two-dimensional framework for HCAI (modified from Shneiderman 2020)**

*5.2.1. R&S in AEC AIR*

The construction industry deals with safety problems and technology tools often attempt to alleviate this grand challenge (Tixier, Hallowell et al. 2016). Therefore, addressing safety issues using AI is a mainstream research topic in construction (Adam 2018, Fang, Ding et al. 2018, Wu, Cai et al. 2019). Construction tasks that should be carried out at a high altitude, extended periods of overtime work, or under any hazardous conditions are desired to be delegated to AI-enabled systems or robots to increase efficiency and eliminate accidents (Brosque, Galbally et al. 2020). Systems that use hazard proximity monitoring are among the most common applications of AI for contraction safety. For example, researchers have implemented a system which sends images from trucks, cranes, and other construction machinery to a database via 5G wireless networks and utilizes AI to evaluate the interactions between workers and equipment (Nozaki, Okamoto et al. 2018). In building construction, a dynamic BIM can provide AEC teams with insights for better planning and more efficient design, construction, and operation/maintenance. This can be improved AI where BIM and AI-based software packages use machine learning algorithms to analyze all aspects of a proposed design or an altered design to ensure it does not interfere with other systems in the building, thus enhancing the reliability of design (Rao 2019). AI can also be used to predict safety measures such as injury severity, injury type, body part impacted, and incident type in a construction project (Baker, Hallowell et al. 2020).

However, regardless of whether the ultimate goal is to address a safety challenge or not, AI and robotics should not pose any safety threats to workers. Furthermore, it has been shown that any technologically advanced system must exhibit R&S to be acceptable in construction (Czarnowski, Dąbrowski et al. 2018). For example, human safety while working with intelligent robots is an important consideration with regards to the robotic acceptance in industrial applications since the type of robot hardware is usually rugged, large, or manipulation-enabled and human humans often feel unsafe working around robots (Bartneck, Kulić et al. 2009, Tan, Duan et al. 2009). Figure 7 shows a few examples of using robotics in building construction.

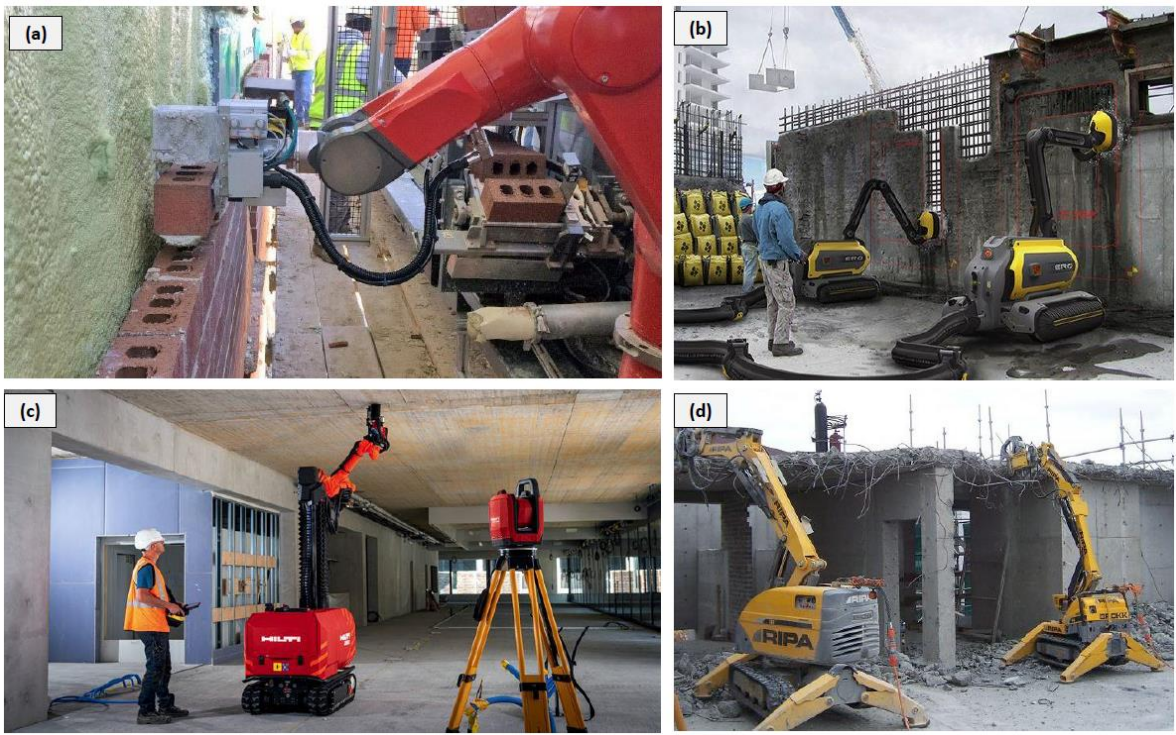

**Figure 7. Examples of application of robotics in building construction: (a) Architecture-related, façade installation (Malewar 2017); (b) Structural-related, shotcrete for shear walls (Building-Design-Construction 2013); (c) Interior layout and MEP related, BIM-**

**enabled mobile ceiling-drilling robot (Hilti-Jaibot); (d) Demolition of buildings and existing structures (Business-Industry-Reports 2018)**

The study of R&S in human-robot interaction in construction can be conducted with two validation approaches; in one approach, robot prototypes operate in physically simulated working environments to identify robot weaknesses in R&S and enhance its functionality. Another approach is to develop predictive models to explicate the perceived R&S of robots and their influence on humans (Hentout, Aouache et al. 2019, Zacharaki, Kostavelis et al. 2020). Models such as the Robot Acceptance Safety Model (RASM) have been developed using the second approach in the construction robotics literature. RASM is a model that integrates Immersive Virtual Environments (IVEs) to analyze the perceived safety of collaborative work between humans and robots. This was done by inviting participants to a laboratory to collaborate and work with a 3D simulated robot regarding a construction-related task using a head-mounted display. According to the IVE results, segregation of human and robot working areas leads to accelerated perceived safety as it facilitates team identification and fosters trusting robots gradually. Moreover, it was deduced that improved safety feeling and comfort of simulation attendees while working with robots signifies their willingness to work with robots in the future (You, Kim et al. 2018). Collaboration and efficiency of human and robot work in industrial settings can be hindered by collisions and other types of accidents (Park, Kim et al. 2007). Active vision-based safety systems have been reported to be one of the best solutions to mitigate collisions and consequently, boost trust in humans to work alongside robots and automation systems. Recent methodologies and advances in vision-based technologies and alarming systems including various methods, sensor types, safety functions, and static/dynamic action of robots to work in their zones have been reviewed and analyzed technically in the literature (Halme, Lanz et al. 2018).

## 5.3. Performance and Robustness (P&R)

Machine competence, which manifests itself in technical performance and ability, is a leading driver of trust in human-machine teaming literature (Lee and See 2004, Hancock, Billings et al. 2011, Schaefer, Chen et al. 2016). To assess the performance of an AI system, measures such as accuracy are commonly used. Accuracy is a gold standard of machine learning models that contributes to the trustworthiness of the AI system (Wing 2020). Also, robust AI developed in one context can be transferred to a wide range of problems within and outside that context in a systematic way manifesting a similar level of performance (Marcus 2020). Theoretically, the two concepts of P&R can be at odds with each other since imperceptible perturbations of the test data (i.e., adversarial examples) may result in incorrect misclassification with high confidence (Dalvi, Domingos et al. 2004, Biggio and Roli 2018, Tsipras, Santurkar et al. 2018, Yang, Rashtchian et al. 2020). However, they are both commonly used to describe the technical reliability of AI systems in the context of trustworthiness. For example, the Ethics Guidelines for Trustworthy AI prepared by the European Commission in 2018 lists robustness defined as "resilience, accuracy, reliability of AI systems" as one of the seven broad key requirements that contribute to trustworthiness (The-European-Commission 2019, Hamon, Junklewitz et al. 2020). Explicit training against known adversarial examples, as well as techniques such as regularization and robust inference are methods to produce robust machine learning models with acceptable accuracy (Madry, Makelov et al. 2017, Zantedeschi, Nicolae et al. 2017). One way that laypeople trust AI models is based on performance metrics such as accuracy. Such accuracy-based trust can be achieved by knowing the stated performance of the system or direct observation or perception of the model's performance in practice. However, if a model's observed accuracy is low, the effect size of the stated accuracy on people's trust is very low, meaning that people's trust in a model is significantly affected by its observed accuracy regardless of its stated accuracy (Yin, Wortman Vaughan et al. 2019).

Regardless of the source of where the performance accuracy comes from, people stop trusting an algorithm as soon as they observe it make a mistake (Dzindolet, Pierce et al. 2002, Dietvorst, Simmons et al. 2015). Laypeople stabilize their trust to a level correlated with the perceived accuracy (Yu, Berkovsky et al. 2016), although system failures have a stronger impact on trust than system successes (Yu, Berkovsky et al. 2017). While P&R have shown to have a significant effect on user's trust in an AI system, there are other machine learning performance metrics such as precision and recall that have been shown to be effective in terms of trust-building. For example, users tend to put more weight on a classifiers' recall rather than its precision when deciding whether the classifiers' performance is acceptable, although the application makes a difference in terms of the weight (Kay, Patel et al. 2015).

*5.3.1. P&R in AEC AI*

Construction use cases are not different from other applications in that the performance of the technology tool plays a key role in trusting and ultimately adopting it. Almost all performance measures in construction ultimately target a metric related to cost or schedule of the project (Cai, Ma et al. 2019, Mohammadpour, Karan et al. 2019, Pan and Zhang 2021). However, immediate gains represented by factors such as quality control results, safety records, sustainability measures, productivity metrics, and inspection outcomes are as important as ultimate cost and schedule benefits in adopting innovations and trusting new technologies in construction (Dulaimi, Nepal et al. 2005, Nitithamyong and Skibniewski 2006, Ishak 2013). Construction is a multi-sector, project-based industry where in any given project, multiple specialty trade contractors are involved. Heterogeneity in projects means that the applications and algorithms of AI and machine learning systems can vary widely from one project type to the other or even within one project (Lattanzi and Miller 2017). Thus, robustness in construction AI translates to the ability to successfully transfer models between different sectors, trades, and even geographic locations considering the uniqueness of any given project without compromising the user's trust in the system's ability. Such P&R must manifest itself in both the technical construction work and the business aspects. In other words, trust cannot be generated if the models have acceptable technical abilities in performing trade work but ignore variables related to business and management processes (Hoch and Brad 2020). For example, a 334 meter-wide dam construction work in Japan uses robots for tasks such as pouring concrete, brushing uneven layers, and dismantling forms (Thornton 2020). Also, within tunnel construction projects, applications such as inspection, maintenance, and health monitoring benefit from AI (Attard, Debono et al. 2018, Leonidas and Xu 2018, Xu and Yang 2020). Robust AI models developed for these tasks in a dam construction project can be effectively transferred and reused in the tunneling project if certain project-specific risk, quality, and inspection considerations are programmed into the design of the model so it results in similar performance acceptable measures.

## 5.4 Privacy and Security (P&S)

P&S are among the important socio-technical considerations pertinent to the use of AIR technologies. Privacy is defined as the right not to be observed and the nature of AI necessitates observing various human-related the phenomenon to learn and improve (Stahl and Wright 2018). Ethical AI is also known as responsible AI and is known as an enabler of engendering trust and scaling AI with confidence (Eitel-Porter 2020). Data security is the most basic and common requirement of responsible AI. Although AI privacy and security are interconnected in most cases, they refer to two different concepts. Security in AI can be associated with ensuring the confidentiality of data, preserving information integrity, and guaranteeing immediate data availability when desired; but privacy is often tied to acquiring, processing, and using personal data. Preventing data breaches by accident and/or improper system engineering and design, ensuring unplanned data corruption, and obstructing outsider's intention to hinder or limit user's access to the system or portal are undoubtedly essential to make sure a designed AI-based system or product cannot be hacked or breached (Wu 2020). The majority of the recent topics regarding ethical AI in the literature are focused on opacity of AI systems, privacy and surveillance, machine ethics (or machine morality), ethical decision making effect of automation on employment, manipulation of behavior, human-robot interaction, bias in decision makings, control of autonomous systems, artificial moral agents, and singularity (Yampolskiy 2018, Müller 2020). In the context of human-robot interaction, P&S have also been identified as a relevant risk to trust (Stuck, Holthausen et al. 2021). Privacy depends on the protection of cybersecurity systems, but is directly related to engendering trust in robots (He, Gray et al. 2020). In the context of human-robot interaction, there are two types of risks to trust from P&S perspective. Privacy situational risk is the belief that a human-robot task or activity will likely expose personal information about the user or their surroundings. Privacy relational risk is the belief that a human-robot task will expose the team or their environment to unauthorized observation or disturbance. Security situational risk is the belief that a human-robot task or situation could make the team vulnerable to crime, sabotage, attack, or some other threat to safety. Security relational risk is the belief that a human-robot could be vulnerable to being misused for crime, sabotage, attack, or some other threat to safety.

### *5.4.1. P&S in AEC AIR*

There are limited studies on the ethical issues and challenges caused by using AIR in the AEC industry. AIR applications require monitoring construction tasks in order to collect data using telecommunications devices, wearable devices (such as VRs, smart hard hat cameras, and sensors), GPS, CCTV cameras, drones, or smartphones (Akhavian and Behzdan 2015, Akhavian and Behzdan 2016, Sakib, Chaspari et al. 2020). While applications vary from worker safety (Nath, Akhavian et al. 2017) to equipment emission monitoring (Akhavian and Behzdan 2013), worker performance and contextual information are involved in one way or the other. Some researchers believe that worker monitoring methodologies that are often used to train machine learning models fail to consider the worker's natural consciousness, intentionality and free will (McAleenan, McAleenan et al. 2018). Another pervasive technology in the AEC domain that enable AIR applications is cloud computing (Bilal, Oyedele et al. 2016). Industrial applications of cloud computing involve P&S challenges for data security, access control, and intrusion prevention (Kim and Park 2020, Hongsong, Yongpeng et al. 2021), but specific solutions for data security in construction projects should be further studied and analyzed (Bilal, Oyedele et al. 2016).

### 6. Future Research

With the observations outlined in the comprehensive overview of the literature on trust in AI presented in Section 5, the authors have identified the following future directions for the AEC research in this area.

### 6.1. Integrated and interdisciplinary studies

There are various lenses through which the identified trust dimensions in this paper can be considered. For example, the T&I concept is sometimes seen as a technical issue and sometimes as a legal or social issue (National Academies of Sciences and Medicine 2018). Dimensions such as R&S can be studied from both the cognitive and emotional perspectives and human trust is generally higher for issues that do not require social or emotional intelligence (Glikson and Woolley 2020). Therefore, it is concluded in this study that regardless of the application area and technology implementation, it is important to see trust in AIR as an interdisciplinary field of research. In AEC, various engineering and management aspects of the projects also affect this conglomerate of fields. Depending on the project phase (e.g., pre-construction, construction, post-construction), project type (e.g., building construction, horizontal construction), application objective (e.g., safety, productivity, sustainability, scheduling), and specific technology being leveraged (e.g., BIM, robotics, mobile computing, blockchain), different trust elements can hinder or drive adoption of AI in AEC projects. A previous study by the authors indicated that regardless of the project phase, type, application, and technology, all the identified trust elements in this study (i.e., T&I, R&S, P&R, and P&S) are key concepts in generating trust in AIR for AEC applications (Emaminejad, North et al. 2021).

### 6.2. Trust calibration

Adjusting human trust levels with the actual abilities of the AIR systems is a key process to ensure overtrust and undertrust are not happening (Muir 1994, Lee and See 2004). Research has shown that creating transparency about an AIR abilities significantly increases the likelihood of building calibrated trust and increasing adoption levels (Wang, Pynadath et al. 2016). However, similar to the dearth of research in trusting and adopting AIR systems in the AEC industry, determining the adequate level of trust that is justified and can be adjusted based on real technical capabilities (i.e., trust calibration) has received very little attention.

### 6.3. Human-centered trust

In AEC, it is vital to promote research that focuses on AIR as a leverage to enable transitioning workers' role to higher-level tasks as opposed to eliminating large segments of the workforce. Research has shown that to initiate trust between humans and AI, AI systems should prove human-centered and a means to serve humankind, upskill workers, and promote human values (Dignum 2017, Lewis, Sycara et al. 2018). Another concept in human-centered research towards trust in AIR that received major attention in fields other than AEC, is the importance of anthropomorphism. Anthropomorphism refers to human-likeness and the perception of technology or an object as having human qualities, such as feelings (Khan and Sutcliffe 2014, Waytz, Heafner et al. 2014).

### 6.4. Enhancing T&I on Value and Mechanism

In order to secure management buy-in and establish trust among end-users, the benefits of adopting AIR should be clearly identified (Son, Park et al. 2012). In addition, the mechanism by which AIR enhances the workflows should be interpreted to those who make adoption decisions as well as the end-users (Legris, Ingham et al. 2003). Addressing these two issues can satisfy the T&I requirement identified in this research. Previous research has demonstrated the significance of presenting benefits in terms of key performance indicators (KPIs) such as time, cost, safety, and sustainability added values (Madaio, Stark et al. 2020). In construction, many studies have shown that perceived added value of new technologies in terms of safety, efficiency, and gains in cost or time saving have a substantial impact on adoption levels of those technologies (FLANAGAN and MARSH 2000, Ahuja, Yang et al. 2010, Habets, Voordijk et al. 2011, Goulding and Alshawi 2012). The important role of workforce training and education in establishing the "what" (i.e., value) and "how" (i.e., mechanism) of introducing new technologies have been also highlighted in the literature (Drobotowicz 2020). Further research can determine what approaches best demonstrate the added values and explain the inner-workings of AIR technology to construction professionals.

### 6.5. Intuitiveness and Familiarity

A recent survey on human-robot collaboration emphasizes the fact that intuitive interfaces drive adoption levels (Chen, Ji et al. 2010). In AEC applications, this translates to human-AI interaction mechanisms that do not entail sophisticated programming or require complex computer skills. For example, advances in natural language processing (NLP) (Hirschberg and Manning 2015), learning from demonstration (LfD) (Argall, Chernova et al. 2009), and active learning (Prince 2004) have emerged in computer science to simplify the ways humans interact with AIR-enabled applications, thus making them easier to adopt. By the same token, a familiar interface results in a more welcoming appearance for the end-user and thus increased levels of trust and adoption (Desai, Medvedev et al. 2012). In AEC applications, it is vital to design AIR technologies that are compatible with the traditional workflows and reinforce best practices as opposed to replacing them. Furthermore, leveraging technologies such as BIM that have become industry standard as a familiar interface can create the intuitiveness and familiarity effects. Future research can explore possibilities that NLP, LfD, active learning, and familiar technologies such as BIM can create to further grow the adoption levels of AIR applications.

### 7. Conclusion

In this paper, a comprehensive review of research on human trust in AIR systems as well as the role of these technologies in the AEC research and industry practice was presented. Trust dimensions were broken down into four main dimensions (i.e., T&I, R&S, P&R, and P&S) to organize the presentation of the review. For each dimension, the literature published after 2010 on two fields of trust in AIR and AIR in AEC, totaling 484 peer-reviewed scholarly papers were thoroughly reviewed. Finally, five directions for the future of research in trust in AIR for AEC applications were proposed. The authors believe that the collection and critical review presented in this paper guide researchers, scholars, and practitioners to reinforce their work within the context of the state-of-the-art.

The presented study has three limitations. While the four dimensions presented in this research are the main trust paradigms identified in the literature, other dimensions such as fairness (Chouldechova and Roth 2018), tangibility (Glikson and Woolley 2020), and availability (Wing 2020) can be included in future reviews. Second, the literature prior to 2010 was not reviewed so a more comprehensive review can also include those publications. Third, the proposed framework for the literature review only considered robots that are powered with AI. A standalone review on trusting robots in AEC may also include robotic systems that are not intelligent or use AI to function.

**Declaration of Competing Interest**

The authors have no competing interest to declare.

**Acknowledgment**

The presented work has been supported by the U.S. National Science Foundation (NSF) CAREER Award through the grant # CMMI 2047138. The authors gratefully acknowledge the support from the NSF. Any opinions, findings, conclusions, and recommendations expressed in this paper are those of the authors and do not necessarily represent those of the NSF.